\documentclass[secnumarabic, graphics,floatfix, nofootinbib,tightenlines,nobibnotes, aps, prl, 12pt]{revtex4-1}
\usepackage{graphicx}
\usepackage[english]{babel}
\usepackage[utf8]{inputenc}
\usepackage{mathrsfs}
\usepackage{amsmath,amssymb}
\usepackage{amsfonts}
\usepackage{multirow}
\usepackage[section]{placeins}

\usepackage{amsfonts}
\usepackage{epstopdf}
\usepackage{latexsym}
\usepackage{amssymb}
\usepackage{amssymb}
\usepackage{color}
\usepackage[center]{subfigure}

\begin{document}

\newcommand{\bq}{\begin{equation}}
\newcommand{\eq}{\end{equation}}
\newcommand{\bqn}{\begin{eqnarray}}
\newcommand{\eqn}{\end{eqnarray}}
\newcommand{\nb}{\nonumber}
\newcommand{\lb}{\label}

\title{Higher dimensional power-Maxwell charged black holes in Einstein and Rastall gravity}

\author{Kai Lin$^{1,2}$}\email{lk314159@hotmail.com}
\author{Yunqi Liu$^{3}$}\email{liuyunqi@hust.edu.cn}
\author{Wei-Liang Qian$^{2,4,3}$}\email{wlqian@usp.br}

\affiliation{$^{1}$ Hubei Subsurface Multi-scale Imaging Key Laboratory, Institute of Geophysics and Geomatics, China University of Geosciences, 430074, Wuhan, Hubei, China}
\affiliation{$^{2}$ Escola de Engenharia de Lorena, Universidade de S\~ao Paulo, 12602-810, Lorena, SP, Brazil}
\affiliation{$^{3}$ Center for Gravitation and Cosmology, College of Physical Science and Technology, Yangzhou University, Yangzhou 225009, China}
\affiliation{$^{4}$ Faculdade de Engenharia de Guaratinguet\'a, Universidade Estadual Paulista, 12516-410, Guaratinguet\'a, SP, Brazil}

\date{April. 27, 2019}

\begin{abstract}
The black hole solutions in higher-dimensional spacetimes with the presence of the power-Maxwell field, surrounded by quintessence, are investigated for Einstein as well as Rastall gravity.
The obtained solutions accommodate for spherical, planar and hyperbolic symmetries with the presence of the cosmological constant.
Besides, we show that several known black hole solutions in literature such as those for linear Maxwell theory and BTZ black hole can be obtained as special cases.
The implications of Rastall's theory related to the present study and the thermodynamics of the black hole solutions are discussed.

\end{abstract}


\maketitle

\section{I Introduction}
\renewcommand{\theequation}{1.\arabic{equation}} \setcounter{equation}{0}

One of the possible generalizations of the general relativity is to relax the constraint that the covariant divergence of the energy-momentum tensor vanishes.
In 1972, Peter Rastall~\cite{agr-rastall-01} proposed that the divergence of the energy-momentum tensor might be nonvanishing and proportional to the gradient of the Ricci scalar.
Rastall's theory assumes that
\bqn
\lb{1}
T^\nu_{\mu;\nu}=a_\mu ,
\eqn
where $a_\mu$ is related to the gradient of the Ricci scalar
\bqn
\lb{2}
T^\nu_{\mu;\nu}=a_\mu=\lambda R_{;\mu} .
\eqn
The corresponding field equation with the presence of the cosmological constant reads 
\bqn
\lb{3}
R_{\mu\nu}-\frac12g_{\mu\nu}R+g_{\mu\nu}\Lambda=\kappa\left(T_{\mu\nu}-\lambda g_{\mu\nu}R\right)
\eqn

The theory effectively implies that the curvature-matter coupling is implemented for modified gravity in a non-minimal fashion.
Owing to such non-minimal coupling, the geometry and matter fields are affected by their mutual changes~\cite{agr-dark-energy-02,agr-dark-energy-03,agr-modified-gravity-general-02,agr-modified-gravity-fR-02,agr-modified-gravity-general-03}.
As a result, the ordinary energy-momentum tensor conservation becomes invalid.
The theory can be viewed as a phenomenological implementation of specific quantum effects in a curved background~\cite{agr-rastall-cosmo-03}.
Besides, in the context of cosmology, the Rastall's theory has been extensively investigated~\cite{agr-rastall-cosmo-01,agr-rastall-cosmo-02,agr-rastall-cosmo-03,agr-rastall-cosmo-04,agr-rastall-cosmo-05,agr-rastall-cosmo-06,agr-rastall-cosmo-07,agr-rastall-cosmo-08,agr-rastall-cosmo-09,agr-rastall-cosmo-10}.
When a free or self-interacting scalar field is included in the Rastall theory~\cite{agr-rastall-04}, it is found that certain exact solutions can be obtained which are associated with $k$-essence theory.
Rastall's theory has been interpreted as a gravitational theory with variable gravitational constant~\cite{agr-rastall-cosmo-01,agr-rastall-cosmo-02}.
In this regard, the standard $\mathrm{\Lambda CDM}$ model is consistently reproduced for the background as well as on linear perturbation level, while additional effects are further explored concerning nonlinear corrections~\cite{agr-rastall-cosmo-03}.

In the vacuum, the only non-trivial static, spherically symmetric solution for the Einstein field equation is the Schwarzschild one.
When an additional scalar field is coupled to the gravity sector, new exact solutions may appear, related to specific Rastall parameter.
Recently, various solutions, and in particular, black hole solutions of the Rastall field equations have been investigated by many authors.
These studies include charged static spherically symmetric black holes~\cite{agr-rastall-05,agr-rastall-04}, Gaussian black holes~\cite{agr-rastall-06,agr-rastall-10}, rotating black holes~\cite{agr-rastall-09,agr-rastall-11}, Abelian-Higgs strings~\cite{agr-rastall-14}, Godel-type black holes~\cite{agr-rastall-13}, 
black branes~\cite{agr-rastall-15}, worm holes~\cite{agr-rastall-07}, neutron stars~\cite{agr-rastall-12}, black hole solutions surrounded by fluid, electromagnetic field~\cite{agr-rastall-02} or quintessence fluid~\cite{agr-rastall-03}, black hole thermodynamics~\cite{agr-rastall-16}, 
higher-dimensional black holes~\cite{agr-rastall-03}, among other theoretical efforts~\cite{agr-rastall-17,agr-rastall-18,agr-rastall-19,agr-rastall-20}.

As Rastall theory has attracted increasing attention recently, besides the fact, as will be addressed below, that it can be viewed as a general formulation of modified gravity.
The present study is devoted to further develop the findings regarding Refs.~\cite{agr-rastall-02,agr-rastall-03,agr-metric-btz-02}.
To be specific, in this work, we aim to investigate the black hole solutions for higher-dimensional spacetimes in Rastall gravity.
The black holes in question are surrounded by quintessence, in the presence of the non-linear electromagnetic field.
General forms of analytic solutions are obtained and presented. 
The paper is organized as follows.
In the following section, we discuss the relationship between Rastall's gravity and other modified theories of gravity briefly. 
In section III, we derive the field equation and present analytic solutions.
Subsequently, the black hole thermodynamics is explored in section IV.
Further discussions and concluding remarks are given in the last section of the paper.

\section{II Remarks on Rastall's gravity}
\renewcommand{\theequation}{2.\arabic{equation}} \setcounter{equation}{0}

Following the line of thought of Rastall's original idea~\cite{agr-rastall-01}, in general, we note that $a_\mu$ can be any vector that vanishes in flat spacetime while satisfying Eq.(\ref{1}).
Though the curvature of the spacetime governs the specific form of $a_\mu$, it is not necessarily given by Eq.(\ref{2}).
In this context, the filed equation becomes 
\bqn
R_{\mu\nu}-\frac{1}{2}g_{\mu\nu}R=\kappa \left(T_{\mu\nu}-{\mathscr A}_{\mu\nu}\right) , \lb{grastall1}
\eqn
with 
\bqn
a_\mu = \nabla^\nu {\mathscr A}_{\nu\mu} \lb{grastall2}.
\eqn 
Such resultant field equation may be referred to as generalized Rastall gravity, which inherits the original spirit of Rastall's theory.
As a matter of fact, many well-known theories of (modified) gravity can be viewed, to a certain degree, as derived by adopting specific forms of ${\mathscr A}_{\nu\mu}$. 
The simplest example is general relativity with the cosmology constant, since one may rewrite the Einstein field equation as 
\bqn
R_{\mu\nu}-\frac{1}{2}g_{\mu\nu}R=\kappa \left(T_{\mu\nu}-\kappa^{-1} g_{\mu\nu}\Lambda\right) .
\eqn
Comparing to Eq.(\ref{grastall1}) and (\ref{grastall2}), one has
\bqn 
{\mathscr A}_{\mu\nu} = \kappa^{-1} g_{\mu\nu}\Lambda
\eqn
and subsequently
\bqn
a_\mu=\nabla^\nu {\mathscr A}_{\nu\mu}=0 .\lb{grastallcondition}
\eqn
Another example is $f(R, T)$ gravity~\cite{agr-modified-gravity-fRT-01}, where  ${\mathscr A}_{\mu\nu}$ is given by
\bqn
{\mathscr A}_{\mu\nu}&=&\kappa^{-1} \left[\frac{\partial f(R,T)}{\partial T}\left(T_{\mu\nu}+g^{\alpha\beta}\frac{\delta T_{\alpha\beta}}{\delta g^{\mu\nu}}\right)+\left(1+\frac{\partial f(R,T)}{\partial R}\right)R_{\mu\nu}\right.\nb\\
&&\left.-\frac{1}{2}g_{\mu\nu}(f(R,T)+R)+\left(g_{\mu\nu}\nabla^2-\nabla_\mu\nabla_\nu\right)\frac{\partial f(R,T)}{\partial R}\right] .
\eqn
if ${\mathscr A}_{\mu\nu}$ satisfies the condition Eq.(\ref{grastallcondition}).

A more sophiscated example is quadratic gravity~\cite{agr-modified-gravity-qg-01}, where the term ${\mathscr A}_{\mu\nu}$ is found to be
\bqn
{\mathscr A}_{\mu\nu}&=&\kappa^{-1} \left[-\frac{1}{2}g_{\mu\nu}\left(\Lambda+\alpha R_{abcd}R^{abcd}+\beta R_{ab}R^{ab}+\gamma R^2\right.\right. \nb\\
&&\left.\left. -(\beta+4\gamma)\nabla^2R\right)+2\alpha R_{acde}R_b^{cde}+2(2\alpha+\beta)R_{acbd}R^{cd}-4\alpha R_{ac}{R^c}_b\right. \\
&&\left.+2\gamma RR_{\mu\nu} +(4\alpha+\beta)\nabla^2R_{\mu\nu}-(2\alpha+\beta+2\gamma)\nabla_\mu \nabla_\nu R\right] . \nb
\eqn
It is straightforward to show that $a_\mu=\nabla^\nu {\mathscr A}_{\nu\mu}$ vanishes in flat spacetimes.
More generally, one may assume the form of ${\mathscr A}$ to be ${\mathscr A}_{\nu\mu}=\lambda g_{\mu\nu}H(R)$, where $H(R)$ is an arbitrary function of the scalar curvature $R$.
However, for this case, it is known that it becomes troublesome to explore the Lagrangian formulation of the theory. 
In this context, the generalized Rastall gravity in terms of the relation $\nabla_\nu T^\nu_\mu=a_\mu=\nabla_\nu {\mathscr A}^\nu_\mu$ can be viewed as a further relaxation of the conservation law of the energy-momentum tensor while satisfying the equivalence principle.

Interestingly, one may also interpret the above discussions from another viewpoint: if the energy-momentum tensor is redefined as $\bar{T}_{\mu\nu}=T_{\mu\nu}-{\mathscr A}_{\mu\nu}$, one subsequently has $\nabla^\mu\bar{T}_{\mu\nu}=0$.
Thus the resulting field equation looks similar to the Einstein's gravity as the conservation of the energy-momentum tensor formally remains unchanged regarding $\bar{T}_{\mu\nu}$~\cite{agr-rastall-05}.
However, as clearly pointed out in Ref.\cite{agr-rastall-18}, one cannot simply define a new energy-momentum tensor and claim the physical content is not affected.
Therefore, as a possible generalization of general relativity, Rastall gravity is an entirely independent theory which possesses prediction power and is ready to accept challenges from observational data.
Last but not least, mathematically, a redefined energy-momentum tensor invalidates the black hole solutions found in Einstein gravity.
Thus the solutions obtained in this work for Rastall gravity is a non-trivial generalization of those for Einstein gravity.

\section{III Charged black hole solutions for $n+1$ dimensional spacetime in Rastall theory}
\renewcommand{\theequation}{3.\arabic{equation}} \setcounter{equation}{0}

For the subject of the present study, let us consider the following static spherically symmetric black hole metric in $n+1$ dimensional spacetime
\bqn
\lb{4}
ds^2=-f(r)dt^2+\frac{dr^2}{f(r)}+r^2d\Sigma^2_{n-1} .
\eqn
For charged black holes, we assume that the electromagnetic field is also static as well as spherically symmetric.
This implies that the only non-zero component is the electric potential, which is given by
\bqn
\lb{5}
A_\mu=\delta^t_\mu A_t(r) .
\eqn
For the angular part, here we define
\bq
\lb{6}
d\Sigma^2_{n-1}= 
\left\{\begin{array}{cc}
d\theta^2+\sin^2\theta d\Omega^2_{n-2}    &  \sigma=1 \cr\\
dx_idx^i                                  &  \sigma=0 \cr\\
d\theta^2+\sinh^2\theta d\Omega^2_{n-2}   &  \sigma=-1
\end{array}\right. ,
\eq
where $d\Omega^2_{n-2}$ is a $n-2$ dimensional unit sphere, and $\sigma = 1, 0, -1$ correspond to solutions with spherical, planar and hyperbolical spacetime symmetries.
It is worth noting that for the $(2+1)$ dimensional spacetime, as $n=2$, the unit sphere $d\Omega^2$ becomes ill-defined, and therefore Eq.(\ref{6}) is only meaningful for the case of $\sigma=0$.

As a generalization, we also consider the action for the electromagnetic field to have the following nonlinear form~\cite{agr-thermodynamics-02,agr-modified-gravity-Lovelock-02,agr-metric-btz-02,agr-modified-gravity-GB-02}
\bqn
\lb{7}
{\cal L}_F=-\left(-\xi{\cal F}\right)^s
\eqn
where ${\cal F}=F_{\mu\nu}F^{\mu\nu}$ and $F_{\mu\nu}=A_{\mu;\nu}-A_{\nu;\mu}$ is the Faraday tensor. 
$\xi$ and $s$ are constants.
One can always choose $\xi>0$ by redefining $F_{\mu\nu}$. 
For $s=1$, the linear Maxwell field is restored.
Also, as shown below, the BTZ-like Maxwell field corresponds to the case where $s=n/2$. 
Subsequently, the field equation is given by
\bqn
\lb{8}
\partial_\mu\left(\sqrt{-g}{\cal F}^{s-1}F^{\mu\nu}\right) = 0 .\label{efmunu}
\eqn
It is straightforward to show that, the following form for the only non-vanishing components of $F^{\mu\nu}$ indeed satisfy Eq.(\ref{efmunu}):
\bqn
\lb{8}
F^{tr}=-F^{rt}=Q\left(2sr-r\right)^{\frac{1-n}{2s-1}} ,
\eqn
and by integrating in $r$, it gives
\bq
\lb{9}
A_t(r)= 
\left\{\begin{array}{lr}
A_0+\frac{Q}{2s-n}(2sr-r)^{1-\frac{n-1}{2s-1}}    &  s\not=\frac{n}{2} \cr\\
A_0+\frac{Q}{n-1}\ln r   &  s=\frac{n}{2}
\end{array}\right. ,
\eq
where $Q$ represents the electric charge of the black hole, which is justified by comparing with the case where $n=3$ and $s=1$.
For the charged black hole, we choose $A_0=0$.
Subsequently, the energy momentum tensor of electromagnetic field is found to be
\bq
\lb{10}
{{E}_\mu}^\nu=-(-\xi)^s\left({\cal F}\right)^{s-1}\left(2sF_{\sigma\mu}F^{\sigma\nu}-\frac{1}{2}\delta^\nu_\mu{\cal F}\right) .
\eq
Here both the electromagnetic and quintessence fields are to be determined dynamically through their roles in the energy-momentum tensor in the field equation.
The total energy-momentum tensor consists of those of the electromagnetic as well as the quintessence field surrounding the black hole
\bq
{T}_{\mu\nu}={E}_{\mu\nu}+T^*_{\mu\nu} ,
\eq
where the latter is characterized by its barotropic equation of state $p = \omega \rho$, subsequently it reads~\cite{agr-rastall-03} 
\bqn
\lb{11}
{T^*}^{t}_t&=&{T^*}^r_r=-\rho(r)\nb\\
{T^*}^{\theta_1}_{\theta_1}&=&{T^*}^{\theta_2}_{\theta_2}=\frac{1}{n-1}\rho(r)\left(n\omega+1\right)
\eqn
Here, for the quintessence field, $\omega=p/\rho$ is found to be $\sim -1$.
Precise measurements for $\omega$ have been carried out by various collaborations.
To be specific, if $\omega$ is assumed to be time-independent, it is found that $\omega=-1.023\pm0.09 ({\text{stat}})\pm0.054 ({\text{syst}})$ by SNLS and SDSS Collaborations, and $\omega=-1.07\pm0.09 ({\text{stat}, 1\sigma})\pm0.13 ({\text{syst}})$ by ESSENCE Collaboration~\cite{book-weinberg-cosmology}.
In our model, it will become clear below that the corrections owing to Rastall's theory take place only when $\omega\not=-1$ (see Eqs.(\ref{21}) and (\ref{23}) below). 

Substituting the energy-momentum tensor as well as the metric into the Rastall field equation Eq.(\ref{2}-\ref{3}), through a somewhat tedious process, one may rewrite the $(tt)$ and $(\theta\theta)$ components of the field equation as the following
\bqn
\lb{fieldEq0022}
0&=&\sigma(n-2)(n-1)-(-2)^s Q^{2s}r^2 (r(2s-1))^{\frac{2 (n-1) s}{1-2 s}}(-\xi)^s+(-1)^s 2^{s+1}Q^{2s}r^2s(r(2s+1))^{\frac{2 (n-1) s}{1-2 s}}(-\xi)^s \nb\\
 &-&2\sigma(n-2)(n-1)\kappa\lambda-2r^2\Lambda+(n-2)(n-1)(2\kappa\lambda-1)f(r)-2r^2\kappa\rho(r)\nb\\
 &+&(n-1)r(4\kappa\lambda-1)f'(r)+2r^2\kappa\lambda f''(r) ,\nb\\
0&=&-(n-2)\left(-\sigma(n-3)(n-2)+(-2)^sQ^{2s}r^2 (r(2s-1))^{\frac{2 (n-1) s}{1-2 s}}(-\xi)^s+2\sigma(n-2)(n-1)\kappa\lambda+2r^2\Lambda\right) \nb\\
&+&\left(-(n-3)(n-2)(n-1)+2(n-2)(n-1)^2\kappa\lambda \right) f(r)+2r^2\kappa (1+n\omega)\rho(r)\nb\\
&+&r\left(-2(r-2)(r-1)+4(n-1)^2\kappa\lambda\right) f'(r)+(n-1)r^2(2\kappa\lambda-1)f''(r) .
\eqn
To proceed, one notices that the ordinary differential equations for $f(r)$ and $\rho(r)$ can be decoupled, and their solutions are therefore obtained.

For $s\not=n/2$, one finds
\bqn
\lb{12}
\rho(r)&=&C_a r^{-\frac{n (\omega +1) (2 \kappa  \lambda  (n+1)-n+1)}{n (2 \kappa  \lambda  (\omega +1)-1)+1}}-\frac{\rho_A(r)}{\rho_B(r)}\nb\\
f(r)&=&\sigma-\frac{2 M_a}{r^{n-2}}+\frac{2 \Lambda  r^2}{n (2 \kappa  \lambda (n+1)-n+1)}-\frac{f_A(r)}{f_B(r)}+\frac{f_C(r)}{f_D(r)} 
\eqn
where
\bqn
\lb{13}
\rho_A(r)&=&\lambda  (n-1) 2^{s+1} e^{2 i \pi  s} s (-n+4 s-1) \times Q^{2 s} \xi ^s (r (2 s-1))^{\frac{2 (n-1) s}{1-2 s}}\nb\\
\rho_B(r)&=&n ((\omega +1) (2 \kappa  \lambda +1)+n (2 \kappa  \lambda  (\omega +1)+2 s \omega -\omega -1)-2 s (4 \kappa  \lambda  (\omega +1)+\omega -1))-2 s\nb\\
f_A(r)&=&2 C_a \kappa  (n (2 \kappa  \lambda  (\omega +1)-1)+1)^2 \times r^{-\frac{(n-1) (n (\omega +1) (2 \kappa  \lambda -1)+2)}{n (2 \kappa  \lambda  (\omega +1)-1)+1}}\nb\\
f_B(r)&=&(n-1) n (2 \kappa  \lambda  (n+1)-n+1) \times((n-1) \omega -2 \kappa  \lambda  (\omega +1))\\
f_C(r)&=&2^s (1-2 s)^2 Q^{2 s} \xi ^s r^{\frac{2 (n-3) s+2}{1-2 s}} (n ((2s-1) \omega-1)+2 s) e^{2 i \pi  s+\frac{2 (n-1) s (\log (r)-\log (r (2 s-1)))}{2 s-1}}\nb\\
f_D(r)&=&(n-2 s) \left\{n^2 [-2 \kappa  \lambda  (\omega +1)-2 s \omega +\omega +1]+n [(\omega +1) (-2 \kappa  \lambda -1)\right.\nb\\
&&\left.+2 s (4 \kappa  \lambda  (\omega +1)+\omega -1)]+2 s\right\}\nb
\eqn

On the other hand, as discussed below, for $s=n/2$, one obtains the BTZ-like solutions as follows
\bqn
\lb{14}
\rho(r)&=&
C_b r^{-\frac{n (\omega +1) (2 \kappa  \lambda  (n+1)-n+1)}{n (2 \kappa  \lambda  (\omega +1)-1)+1}}+\frac{(-1)^{n}\lambda   2^{n/2} (n-1) \xi ^{n/2} Q^n }{[2 \kappa  \lambda  (\omega +1)-n \omega +\omega](nr-r)^{n} }\nb\\
f(r)&=&\sigma-\frac{2 M_b} {r^{n-2}}+\frac{2 \Lambda  r^2}{n (2 \kappa  \lambda  (n+1)-n+1)}-\frac{(-1)^{n} 2^{n/2} r \omega  \xi ^{n/2} Q^n (nr-r)^{1-n}\ln (r)}{2 \kappa  \lambda  (\omega +1)-n \omega +\omega }\nb\\
&&-\frac{2 C_b \kappa  (n (2 \kappa  \lambda  (\omega +1)-1)+1)^2}{(n-1) n (2 \kappa  \lambda  (n+1)-n+1)} \times\frac{r^{-\frac{(n-1) (n (\omega +1) (2 \kappa  \lambda -1)+2)}{n (2 \kappa  \lambda  (\omega+1)-1)+1}}}{(n-1) \omega -2 \kappa  \lambda  (\omega +1)}
\eqn
where $M_a$, $M_b$, $C_a$ and $C_b$ are constants of integration, and their substantial physical interpretations will be studied in the next section. 
The black hole solutions in Einstein gravity are obtained by taking $\lambda=0$ in the above expressions.

It is meaningful to express the above solutions regarding constants with straightforward physical content.
This will be carried out in the following section.
In what follows, we will use the subscript $a$ to indicate solutions for the $s\not=n/2$ case, and $b$ to represent those for the $s=n/2$ case.

\section{IV Black hole thermodynamics}
\renewcommand{\theequation}{4.\arabic{equation}} \setcounter{equation}{0}

Black hole thermodynamics require that the mass, temperature, entropy, electric charge and electrostatic potential of the static black hole satisfy the first law
\bqn
\lb{15}
d{\cal M}=T_hdS_h+\Phi_h dQ
\eqn
where ${\cal M}$ and $\Phi$ are the mass and electrostatic potential of the black hole, respectively. 
In this section, we focus on the property of black hole thermodynamics at event horizon, where the entropy and temperature for a static black hole are given by
\bqn
\lb{16}
S_h=\frac{A(r_0)}{4},~~~~T_h=\left.\frac{f'(r)}{4\pi}\right|_{r=r_0}
\eqn
where $r_0$ is the radius of the event horizon which satisfies $f(r_0)=0$, and $A(r_0)$ is the area of the event horizon in $n+1$ dimensional spacetime, which is given in terms of the $\Gamma$ function
\bqn
\lb{17}
A_h=2\pi^{n/2}r_0^{n-1}/\Gamma\left(\frac{n}{2}\right)
\eqn
Eq.(\ref{15}) implies
\bqn
\lb{18}
T_h=\left.\frac{\partial{\cal M}}{\partial S}\right|_{\Phi},~~~\Phi_h=\left.\frac{\partial{\cal M}}{\partial Q}\right|_{T_h}
\eqn
which can be used to obtain the masses of the black holes.
In fact, it is not difficult to show that the black hole mass $\cal M$ is proportional to the constant of integration $M$.
To be more specific, one finds that the proportionality constant to be $\frac{n-1}{4\pi^{1-\frac{n}{2}}\Gamma\left(\frac{n}{2}\right)}$, namely, 
\bqn
\lb{19}
{\cal M}=
\left\{\begin{array}{cc}
\frac{n-1}{4\pi^{1-\frac{n}{2}}\Gamma\left(\frac{n}{2}\right)}M_a & x\ne \frac n2 \cr\\
\frac{n-1}{4\pi^{1-\frac{n}{2}}\Gamma\left(\frac{n}{2}\right)}M_b & x=\frac n2 
\end{array}\right. .
\eqn
Moreover, the electrostatic potentials are found to be
\bqn
\lb{20}
\Phi=
\left\{\begin{array}{cc}
\frac{(n-1) \pi ^{\frac{n}{2}-1} 2^{s-2} e^{2 i \pi  s} s (2 s-1)^{2-\frac{2 (n-1) s}{2 s-1}} \varphi_a}{(2 s-n) \Gamma \left(\frac{n}{2}\right) \varrho_a} & x\ne \frac n2 \cr\\
-\frac{(-1)^{n} 2^{\frac{n}{2}-3} (n-1)^{2-n} n \pi ^{\frac{n}{2}-1} \omega  \xi ^{n/2} Q^{n-1} \ln r_0}{\Gamma \left(\frac{n}{2}\right) (2 \kappa  \lambda  (\omega+1)-n \omega +\omega )} & x=\frac n2 
\end{array}\right. ,
\eqn
where
\bqn
\varphi_a&=&Q^{2 s-1} \xi ^s r_0^{\frac{n-2 s}{1-2 s}} (n ((2 s-1) \omega -1)+2 s)\nb\\
\varrho_a&=&n^2 [2 \kappa  \lambda  (\omega +1)+2 s \omega -\omega -1]+n [(\omega +1) (2 \kappa  \lambda +1)\\
&&-2 s (4 \kappa  \lambda  (\omega +1)+\omega -1)]-2 s \nb.
\eqn
Let's compare the above form of electrostatic potentials with Eq.(\ref{9}) evaluated at event horizon $r_0$, one finds that the resultant relation can be used to further constrain the model parameters.
As mentioned by the end of the previous section, we now proceed to rewrite the black hole solutions in terms of physical quantities such as the electric charge $Q$, black hole mass ${\cal M}$ and the radius of event horizon $r_0$.
Besides, one also evaluates quantities such as $\Phi_h$, $S_h$, and $T_h$ in terms of the above variables.

For $s\not=n/2$, one finds
\bqn
\lb{21}
\rho_a&=&\frac{C_a}{r^{r_A}}-\frac{8 \lambda  \pi ^{1-\frac{n}{2}} Q^2 (n-4 s+1) (2 s-1)^{\frac{n-2 s}{1-2 s}-2} \Gamma \left(\frac{n}{2}\right) }{[n ((2 s-1) \omega-1)+2 s] r^{\frac{2 (n-1) s}{2 s-1}}}\nb\\
f_a&=&\sigma-\frac{8\Gamma\left(\frac{n}{2}\right){\cal M}_a}{\pi ^{\frac{n}{2}-1}(n-1)r^{n-2}}+\frac{2 \Lambda  r^2}{n (2 \kappa  \lambda  (n+1)-n+1)}+\frac{4  Q^2 \Gamma \left(\frac{n}{2}\right)\left(2 s-1\right){}^{\frac{n-2 s}{1-2 s}} r^{\frac{2 (n-3) s+2}{1-2 s}}}{\pi ^{\frac{n}{2}-1}(n-1) s (n-2 s)}\nb\\
&&-\frac{2 \kappa  C_a (n (2 \kappa  \lambda  (\omega +1)-1)+1)^2 }{(n-1) n (2\kappa  \lambda  (n+1)-n+1) }\times\frac{r^{-r_B}}{(n-1) \omega -2 \kappa  \lambda  (\omega +1)}\nb\\
\Phi_a&=&-\frac{Q \left(r_0 (2 s-1)\right){}^{\frac{n-2 s}{1-2 s}}}{2 s-n}\\
S_a&=&\frac{\pi ^{n/2} r_0^{n-1}}{2 \Gamma \left(\frac{n}{2}\right)}\nb\\
T_h&=&\frac{\Lambda  r_0}{\pi  n (2 \kappa  \lambda  (n+1)-n+1)}+\frac{2 (n-2)\Gamma\left(\frac{n}{2}\right){\cal M}_a}{(n-1)\pi ^{n/2} r_0^{n-1}}-\frac{2  Q^2  ((n-3) s+1) \Gamma \left(\frac{n}{2}\right) \left(r_0 (2 s-1)\right){}^{\frac{1-n}{2 s-1}}}{\pi ^{n/2}(n-1) s (n-2s)r_0^{n-2}}\nb\\
&&+\frac{\kappa  C_a  r_0^{r_C}}{2 \pi  n (2 \kappa  \lambda  (n+1)-n+1) }\times\frac{(n (\omega +1) (2 \kappa  \lambda -1)+2) (n (2 \kappa  \lambda  (\omega +1)-1)+1)}{(n-1) \omega -2 \kappa  \lambda  (\omega +1)}\nb
\eqn
with
\bqn
\lb{22}
r_A&=&\frac{n (\omega +1) (2 \kappa  \lambda  (n+1)-n+1)}{n (2 \kappa  \lambda  (\omega +1)-1)+1}\nb\\
r_B&=&\frac{(n-1) (n (\omega +1) (2 \kappa  \lambda -1)+2)}{n (2 \kappa  \lambda  (\omega +1)-1)+1}\nb\\
r_C&=&-\frac{(n-1) (n (\omega +1) (2 \kappa  \lambda -1)+2)}{n (2 \kappa \lambda  (\omega +1)-1)+1}-1
\eqn

On the other hand, for BTZ-like solutions with $s=n/2$, we have
\bqn
\lb{23}
\rho_b&=&\frac{C_b} {r^{r_A}}+\frac{8 \lambda   Q^2  \Gamma\left(\frac{n}{2}\right) \ln \left(nr_0-r_0\right)}{r^{n}\pi ^{\frac{n}{2}-1}(n-1)^2 n \omega  \ln r_0}\nb\\
f_b&=&\sigma-\frac{8  \Gamma \left(\frac{n}{2}\right){\cal M}_b}{\pi ^{\frac{n}{2}-1}(n-1)r^{n-2}}+\frac{2 \Lambda  r^2}{n (2 \kappa  \lambda (n+1)-n+1)}-\frac{8 \pi ^{1-\frac{n}{2}} Q^2 r^{2-n} \Gamma \left(\frac{n}{2}\right) \ln (r) \ln \left(nr_0-r_0\right)}{(n-1)^2 n \ln \left(r_0\right)}\nb\\
&&-\frac{2 \kappa  C_b (n (2 \kappa  \lambda  (\omega +1)-1)+1)^2 }{(n-1) n (2\kappa  \lambda  (n+1)-n+1)}\times\frac{r^{-r_B}}{ (n-1) \omega -2 \kappa  \lambda  (\omega +1)}\nb\\
\Phi_b&=&-\frac{Q \ln \left((n-1) r_0\right)}{n-1}\\
S_b&=&\frac{\pi ^{n/2} r_0^{n-1}}{2 \Gamma \left(\frac{n}{2}\right)}\nb\\
T_h&=&\frac{\Lambda  r_0}{\pi  n (2 \kappa  \lambda  (n+1)-n+1)}+\frac{2 (n-2)  {\cal M}_b  \Gamma\left(\frac{n}{2}\right)}{\pi ^{n/2}(n-1)r_0^{n-1}}+\frac{2 Q^2 \Gamma \left(\frac{n}{2}\right) \ln\left((n-1) r_0\right)\left(\ln\left(r_0^{n-2}\right)-1\right)}{ \pi ^{n/2}(n-1)^2 n \ln\left(r_0\right)r_0^{n-1} }\nb\\
&&+\frac{\kappa  C_b  r_0^{r_C}}{2 \pi  n (2 \kappa  \lambda  (n+1)-n+1)}\times\frac{(n (\omega +1) (2 \kappa  \lambda -1)+2) (n (2 \kappa  \lambda  (\omega +1)-1)+1)}{ (n-1) \omega -2 \kappa  \lambda  (\omega +1)} . \nb
\eqn
We note that both ${\cal M}_a$ and ${\cal M}_b$ can be considered as the functions of $r_0$, which are determined by $f_a(r=r_0)=0$ and $f_b(r=r_0)=0$ respectively.

Besides, if the spacetime possesses a cosmological horizon, it implies that $f$ can be written as 
\bqn
f(r)=H(r)(r-r_0)(r_c-r) ,
\eqn
with $H(r) > 0$ for $r_0<r<r_c$. 
The area law of entropy holds at cosmological horizon~\cite{agr-thermodynamics-01,agr-thermodynamics-02}, which gives
\bqn
S_c=\frac{\pi^{n/2}r_c^{n-1}}{2\Gamma\left(\frac{n}{2}\right)} .
\eqn

In addition, we evaluate the heat capacity related to the event horizon.
The definition of heat capacity reads
\bqn
C=\frac{\partial {\cal M}}{\partial T}.
\eqn
In what follows, we only focus on three, four, and five dimensional spherically symmetric cases, which corresponds to $\sigma=1$, and we also take $\kappa\lambda=1$ without loss of generality.

In three dimensional spacetime, for both the cases $s\not=1$ and $s=1$ (the BTZ black hole), the heat capacities are respectively given by
\bqn
C_{3a}&=&\frac{Q^2 \left(r_0 (2 s-1)\right){}^{\frac{1}{1-2 s}} (-6 s \omega -7 s+5 \omega +5)}{\pi  (s-1) s (4 \omega +3)}\nb\\
&&+\frac{\Lambda  r_0^2 \omega +\Lambda  r_0^2+\omega +2}{8 \pi  r_0 \omega +6 \pi  r_0}-\frac{(4 \omega +8) \mathcal{M}}{4 \pi  r_0 \omega +3 \pi  r_0},\nb\\
C_{3b}&=&-\frac{Q^2 \left(2 \omega  \log \left(r_0\right)+4 \log \left(r_0\right)+4 \omega +3\right)}{4 \pi  r_0 \omega +3 \pi  r_0}\nb\\
&&+\frac{\Lambda  r_0^2 \omega +\Lambda  r_0^2+\omega+2}{8 \pi  r_0 \omega +6 \pi  r_0}-\frac{(4 \omega +8) \mathcal{M}}{4 \pi  r_0 \omega +3 \pi  r_0}.
\eqn
We note that the heat capacities is positive when the black hole mass $\cal M$ is larger than a critical value ${\cal M}_c$.
However, the heat capacity becomes negative as the mass is smaller than ${\cal M}_c$, as for other well-known black hole metrics.
For $s\not=1$, the value of the critical mass reads
\bqn
{\cal M}_{c}&=&\big[10 Q^2 r_0 (\omega +1) \left(r_0 (2 s-1)\right){}^{\frac{1}{1-2 s}}\nb\\
&&-s \left(2 Q^2 r_0 (6 \omega +7) \left(r_0 (2 s-1)\right){}^{\frac{1}{1-2 s}}+\Lambda  r_0^2 (\omega +1)+\omega +2\right)\nb\\
&&+s^2 \left(\Lambda  r_0^2 (\omega +1)+\omega +2\right)\big]/\left[8 (s-1) s (\omega +2)\right] ,
\eqn
while for the BTZ case, we have
\bqn
{\cal M}_{c}=\frac{-4 Q^2 \omega  \log \left(r_0\right)-8 Q^2 \log \left(r_0\right)-8 Q^2 \omega -6 Q^2+\Lambda  r_0^2 \omega +\Lambda  r_0^2+\omega +2}{8 \omega +16}.
\eqn

In four dimensional spacetime, for the $s\not=3/2$ case
\bqn
C_{4a}&=&\frac{Q^2 \left(r_0 (2 s-1)\right){}^{\frac{2}{1-2 s}} (9 (\omega +1)-2 s (3 \omega +5))}{4 \pi  r_0 s (2 s-3) (3 \omega +2)}+\frac{\Lambda  r_0^2 \omega +\Lambda  r_0^2+3 \omega +5}{12 \pi  r_0 \omega +8 \pi  r_0}-\frac{3 \mathcal{M}}{6 \pi  r_0^2 \omega +4 \pi  r_0^2}\nb\\
{\cal M}_{c}&=&\big[9 Q^2 r_0 \omega  \left(r_0 (2 s-1)\right){}^{\frac{2}{1-2 s}}-6 Q^2 r_0 s \omega  \left(r_0 (2 s-1)\right){}^{\frac{2}{1-2 s}}+9 Q^2 r_0 \left(r_0 (2  s-1)\right){}^{\frac{2}{1-2 s}}\nb\\
&&-10 Q^2 r_0 s \left(r_0 (2 s-1)\right){}^{\frac{2}{1-2 s}}+2 \Lambda  r_0^3 s^2 \omega +2 \Lambda  r_0^3 s^2+6 r_0 s^2 \omega +10 r_0 s^2-3 \Lambda  r_0^3 s \omega\nb\\
&& -3 \Lambda  r_0^3 s-9 r_0 s \omega -15 r_0 s\big/(12s^2-18s) ,
\eqn
and for the $s=3/2$ case
\bqn
C_{4b}&=&-\frac{3 Q^2 \log \left(r_0\right)+Q^2 (3 \omega +2+\log (8))-3 \left(\Lambda  r_0^3 (\omega +1)+r_0 (3 \omega +5)-6 \mathcal{M}\right)}{12 \pi  r_0^2 (3 \omega +2)}\nb\\
{\cal M}_{c}&=&\frac{1}{18} \left(-3 Q^2 \log \left(r_0\right)-3 Q^2 \omega -2 Q^2-3 Q^2 \log (2)+3 \Lambda  r_0^3 \omega +3 \Lambda 
   r_0^3+9 r_0 \omega +15 r_0\right).\nb\\
\eqn

In five dimensional spacetime, for the $s\not=2$ case
\bqn
C_{5a}&=&\big[16 (s-2) s (\omega -2) \mathcal{M}-2 Q^2 r_0 \left(r_0 (2 s-1)\right){}^{\frac{3}{1-2 s}} (s (4 \omega +13)-14 (\omega +1))\nb\\
&&+3 \pi  r_0^2 (s-2) s\left(\Lambda  r_0^2 (\omega +1)+6 \omega +9\right)\big]/\left[6 \pi ^2 r_0^3 (s-2) s (8 \omega +5)\right]\nb\\
{\cal M}_{c}&=&\frac{2 Q^2 \left(r_0 (2 s-1)\right){}^{\frac{3}{1-2 s}} (s (4 \omega +13)-14 (\omega +1))-3 \pi  r_0 (s-2) s \left(\Lambda  r_0^2 (\omega +1)+6 \omega+9\right)}{16 (s-2) s (\omega -2)r_0^{-1}} ,\nb\\
\eqn
and for the $s=2$ case
\bqn
C_{5b}&=&\frac{4 Q^2 (\omega -2) \log \left(3 r_0\right)+Q^2 (-(8 \omega +5))+9 \pi  r_0^2 \left(\Lambda  r_0^2 (\omega +1)+6 \omega
   +9\right)+48 (\omega -2) \mathcal{M}}{18 \pi ^2 r_0^3 (8 \omega +5)}\nb\\
{\cal M}_{c}&=&\frac{-4 Q^2 (\omega -2) \log \left(3 r_0\right)+Q^2 (8 \omega +5)-9 \pi  r_0^2 \left(\Lambda  r_0^2 (\omega +1)+6 \omega +9\right)}{48 (\omega -2)} .
\eqn

\section{V Discussions and concluding remarks}
\renewcommand{\theequation}{5.\arabic{equation}} \setcounter{equation}{0}

Now, let us discuss a few known black hole solutions which can be obtained as special cases of the solution found in this work, Eq.(\ref{12}) and Eq.(\ref{14}).
First, we note that owing to Eq.(\ref{8}), $s = n/2$ is associated with a distinct class of solutions.
In particular, for the (2+1)-dimensional spacetime with linear Maxwell theory, namely, $n=2, s=1$, the BTZ black hole solution is restored.
Therefore the solutions satisfying $s = n/2$ are referred to in the text as BTZ-like solutions.
In $(3+1)$ asymptotic spacetime, the black hole solutions with linear Maxwell theory have been studied in Ref.~\cite{agr-rastall-05} where the surrounding fluid satisfies the barotropic equation of state, $p=\omega \rho$.
On the other hand, when the Maxwell field is removed, the black hole solution in higher dimensional spacetime with the presence of quintessence fluid has been investigated in~\cite{agr-rastall-03} recently.
It is readily to verify that the solutions presented in~\cite{agr-rastall-03} are restored by taking $s=0$ in Eq.(\ref{12}).
If one further considers the case where $n=3$, the results in~\cite{agr-rastall-02} are reproduced.
Also, the solution found in~\cite{agr-hawking-radiation-02} is obtained in the limit $\kappa \lambda\rightarrow 0$.
Moreover, in $(2+1)$ dimensional spacetime, only the planar black hole solutions persist, which corresponding to take $\sigma =0$ in the general solution Eq.(\ref{12}) and Eq.(\ref{14}).
Last but not least, the present results restore to the charged BTZ-like solutions in higher dimensions~\cite{agr-metric-btz-02} by taking $\lambda, C_a, C_b \to 0$.

To summarize, in the present work, explicit forms of the black hole solutions in Einstein and Rastall gravity are obtained, where one considers the power-Maxwell field in an arbitrary $n$ dimensional spacetime, surrounded by quintessence fluid.
The solutions with spherical, planar, as well as hyperbolical spacetime symmetries are obtained and expressed in a compact form corresponding to $\sigma = 1, 0, -1$.
In particular, a class of BTZ-like black hole solutions is found, which is associated with the BTZ black hole in (2+1) dimensional anti-de Sitter spacetime, as well as with other BTZ-like solutions in higher dimensions. 
In regards to the black hole thermodynamics, we present the thermodynamic quantities regarding those related to the hair of the black hole.
We found the constant of integration, $M$, is proportional to the mass of the black hole, $\cal{M}$, and therefore carries the physical content of the latter.
The present study may pave the way for further investigation on quasinormal modes as well as Hawking radiation of the black hole solutions, especially for the case of higher dimensional spacetimes, which is relevant to many related topics. 

\section*{Acknowledgements}
We are thankful for valuable discussions with Alan B. Pavan. 
We gratefully acknowledge the financial support from Brazilian funding agencies Funda\c{c}\~ao de Amparo \`a Pesquisa do Estado de S\~ao Paulo (FAPESP), 
Conselho Nacional de Desenvolvimento Cient\'{\i}fico e Tecnol\'ogico (CNPq), Coordena\c{c}\~ao de Aperfei\c{c}oamento de Pessoal de N\'ivel Superior (CAPES), 
and National Natural Science Foundation of China (NNSFC) under contract No.11805166.


\bibliographystyle{h-physrev}
\bibliography{references_qian}

\end{document}